# Berezinskii-Kosterlitz-Thouless Crossover in a Trapped Atomic Gas

Zoran Hadzibabic, Peter Krüger, Marc Cheneau, Baptiste Battelier & Jean Dalibard

*Laboratoire Kastler Brossel, Ecole normale supérieure, 24 rue Lhomond, F-75231 Paris CEDEX 05, France*

**Any state of matter is classified according to its order, and the kind of order a physical system can posses is profoundly affected by its dimensionality. Conventional long-range order, like in a ferromagnet or a crystal, is common in three-dimensional (3D) systems at low temperature. However, in two-dimensional (2D) systems with a continuous symmetry, true long-range order is destroyed by thermal fluctuations at any finite temperature[1,2]. Consequently, in contrast to the 3D case, a uniform 2D fluid of identical bosons cannot undergo Bose-Einstein condensation. Nevertheless, it can form a "quasi-condensate" and become superfluid below a finite critical temperature. The Berezinskii-Kosterlitz-Thouless (BKT) theory[3,4] associates this phase transition with the emergence of a topological order, resulting from the pairing of vortices with opposite circulations. Above the critical temperature, proliferation of unbound vortices is expected. Here we report the observation of a BKT-type crossover in a trapped quantum degenerate gas of rubidium atoms. Using a matter wave heterodyning technique, we observe both the long-wavelength fluctuations of the quasi-condensate phase and the free vortices. At low temperatures, the gas is quasi-coherent on the length scale set by the system size. As the temperature is increased, the loss of long-range coherence coincides with the onset of proliferation of free vortices. Our results provide direct experimental evidence for the microscopic mechanism underlying the BKT theory, and raise new questions regarding coherence and superfluidity in mesoscopic systems.**

The BKT mechanism is very different from the usual finite-temperature phase transitions. It does not involve any spontaneous symmetry breaking and emergence of a spatially uniform order parameter. Instead, the low temperature phase is associated with a quasi-long-range order, with the correlations of the order parameter (e.g. the macroscopic wave function of a Bose fluid) decaying algebraically in space. Above the critical temperature this quasi-long-range order is no longer maintained and the correlations decay exponentially. This picture is applicable to a wide variety of 2D phenomena including the superfluidity in liquid helium films[5], the superconducting transition in arrays of Josephson junctions[6], and the collision physics of 2D atomic hydrogen[7]. These experiments have provided evidence for the BKT phase transition by looking at the macroscopic properties of the system, but could not reveal its microscopic origin, i.e. binding and unbinding of vortex-antivortex pairs[3,4].

Harmonically trapped atomic gases generally provide an excellent testing ground for the theories of many-body physics. In particular, they are well suited for the preparation of low dimensional systems and the detection of individual vortices. Quasi-2D quantum degenerate Bose gases have been produced in single "pancake" traps or at the nodes of one-dimensional (1D) optical lattice potentials[8,9,10,11,12,13,14,15]. Recently, matter wave interference between small disc-shaped quasi-condensates has revealed occasional presence of free vortices[16], but a systematic temperature study was not possible. Theoretically, since the density of states in a 2D harmonic trap allows for finite temperature Bose-Einstein condensation in an ideal gas[17], the nature of the superfluid transition in an interacting gas has been a topic of some debate[18,19,20,21,22,23,24]. Our results indicate that the BKT picture is applicable to these systems even though in our finite-size system the transition occurs as a finite-width crossover rather than a sharp phase transition[25].

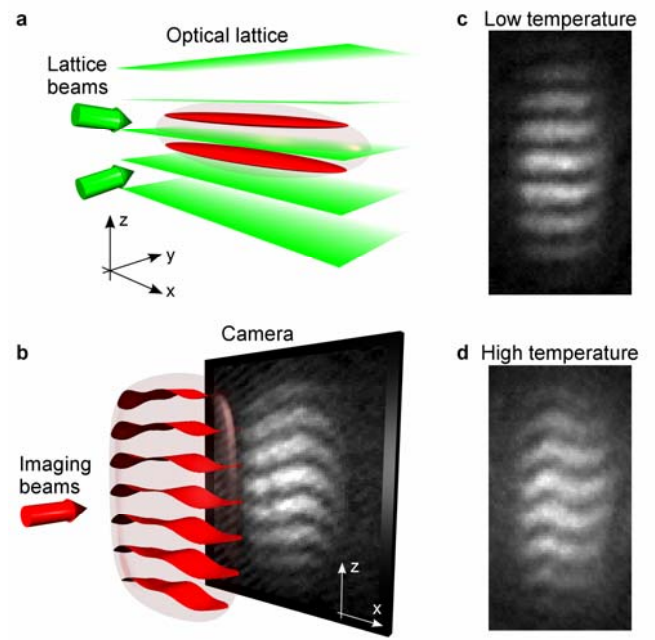

**Figure 1 | Probing the coherence of 2D atomic gases using matter wave heterodyning. a,** An optical lattice potential of period $d = 3$ μm along the vertical direction $z$ is formed by two laser beams with a wavelength of 532 nm intersecting at a small angle. It is used to split a quantum degenerate 3D gas into two independent planar systems. The transparent ellipsoid indicates the shape of the gas before the lattice is ramped up. **b,** After abrupt switching off of the confining potential, the two atomic clouds expand, overlap and interfere. The interference pattern is recorded onto a CCD camera using the absorption of a resonant probe laser. The waviness of the interference fringes contains information about the phase patterns in the two planar systems. **c and d,** Examples of interference patterns obtained at a low and a high temperature, respectively.

We start our experiments with a quantum degenerate 3D cloud of $^{87}$Rb atoms, produced by radio-frequency evaporation in a cylindrically symmetric magnetic trap. Next, a 1D optical lattice with a period of $d = 3$ μm along the vertical direction $z$ is used to split the 3D gas into two independent clouds and to compress them into the 2D regime (Figure 1a). To minimize heating and ensure thermal equilibrium, the lattice potential is ramped up slowly over 500 ms, and the clouds are allowed to equilibrate for another 200 ms. At full laser power, the height of the lattice potential is $V_0 / h = 50$ kHz, where $h$ is Planck's constant. At this lattice height, the tunnelling between the two planes is negligible on the time-scale of the experiment, and the motion along the tight confining direction $z$ is "frozen out" [14,16]. The two clouds form parallel, elongated 2D strips, characterized by the harmonic trapping frequencies of 11 Hz, 130 Hz, and 3.6 kHz along the $x$, $y$, and $z$ directions, respectively. The number of condensed atoms per plane is a function of temperature and varies between 0 and $5\times10^4$, whereas the total atom number per plane is $\sim 10^5$. For the largest (quasi-)condensates, the Thomas-Fermi (TF) approximation yields 120 μm and 10 μm for the $x$ and $y$ lengths of the strips, respectively. The corresponding chemical potential and healing length are $\mu/h = 1.7$ kHz and $\xi = 0.2$ μm.

After the trapped 2D gases have equilibrated, all confining potentials are suddenly turned off. The two clouds expand predominantly perpendicular to the $xy$ plane and, as they overlap, a 3D matter wave interference pattern forms[26]. After $t = 20$ ms of "time-of-flight" (TOF) expansion, the projection of the 3D interference pattern onto the $xz$ plane is recorded on a CCD camera, using a resonant probe laser directed along $y$ (Figure 1b). At any fixed position $x$, the interference pattern along $z$ is characterized by its contrast $c(x)$ and phase $\varphi(x)$. To extract these two parameters we fit the density distribution with a function:

$$F(x,z) = G(x,z)\left[1 + c(x)\cos(2\pi z/D + \varphi(x))\right]$$

where $G(x,z)$ is a gaussian envelope, $D = ht/md$ is the period of the interference fringes, and $m$ is the atomic mass. Intuitively, $c(x)$ is a measure of the local coherence in the 2D clouds (with some coarse grain averaging due to the integration along the imaging axis $y$) while the variation of $\varphi(x)$ with $x$ is a measure of the long-range coherence. With increasing temperature, the presence of phase fluctuations in the two planes increases the waviness of the interference fringes, i.e. the fluctuations in $\varphi(x)$ (c.f. Figure 1c and 1d).

In order to explore different temperature regimes for the 2D gas, we vary the final radio-frequency $\nu_{rf}$ used in the evaporative cooling of the initial 3D gas. The temperature $T^{3D}$ is proportional to $\Delta\nu = \nu_{rf} - \nu_{rf}^{(min)}$, where $\nu_{rf}^{(min)}$ is the final radio-frequency which completely empties the trap. We explore the range between the onset of condensation in the 3D gas ($T^{3D}$=150 nK) and a quasi-pure 3D BEC. As the lattice is ramped up, the temperature of the compressed gas can increase significantly (2-3 times), but precise direct thermometry in the lattice is difficult. Instead, in order to quantify the degeneracy of the 2D system, we measure the local contrast in the center of the interference pattern, $c_0 = \langle c(0) \rangle$, where $\langle ... \rangle$ denotes an average over many images recorded under the same experimental conditions (temperature and atom number).

The dependence of $c_0$ on the initial $T^{3D}$ (i.e. $\Delta\nu$), is shown in Figure 2. The interference fringes are visible for $\Delta\nu < 35$ kHz which closely corresponds to the range of condensation in the initial 3D gas. As $\Delta\nu$ is lowered, $c_0$ grows smoothly. For $\Delta\nu$ below ~12 kHz, the initial 3D BEC is essentially pure and $c_0$ saturates at about 30%. In an ideal experiment, the expected contrast at zero temperature is $c_0$=1. The finite resolution of our imaging system limits the maximal observable contrast to about 60%. We attribute the difference between expected and measured maximal contrasts to the residual heating of the gas in the optical lattice, caused in particular by the 3-body recombination processes. This hypothesis is supported by the fact that the atoms experience the lattice potential during 700 ms, which is not negligible compared to the measured lifetime of 2.5 s for the atom cloud in the lattice. In the following we use $c_0$ rather than $T^{3D}$ as a direct measure of the degeneracy of the 2D gas.

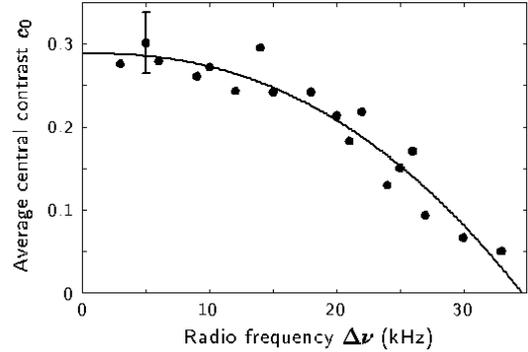

**Figure 2 | Local coherence as a thermometer.** The average central contrast $c_0$ of the interference patterns is plotted as a function of the parameter $\Delta\nu$ controlling the temperature of the 3D gas before loading the optical lattice. The solid line is a fit to the data using the empirical function $c_0 = c_{max}\left[1 - (\Delta\nu/\Delta\nu_0)^\gamma\right]$, with $c_{max} = 0.29$ (±0.2), $\Delta\nu_0 = 35$ (±1) kHz and $\gamma = 2.3$ (±0.4). The total number of images used for the plot is 1200, corresponding to 41 measurements of $c_0$. Different measurements of $c_0$ taken at equal $\Delta\nu$ have been averaged. The displayed error bar indicates the largest standard deviation.

We now turn to a quantitative analysis of long-range correlations as a function of temperature. The coherence in the system is encoded in the first order correlation function:

$$g_1(\vec{r},\vec{r}\,') = \langle \psi^*(\vec{r})\psi(\vec{r}\,') \rangle$$

where $\psi(\vec{r})$ is the fluctuating bosonic field at position $\vec{r}$. From interference signals recorded at different positions along the $x$ axis, one can extract information about $g_1$, as well as higher order correlation functions[27]. Here we adopt a particularly elegant analysis method proposed by Polkovnikov, Altman and Demler[28]. The idea is to partially integrate the 3D interference pattern over lengths $L_x$ and $L_y$, along the $x$ and $y$ directions respectively, and study how the resulting contrast $C$ decays with the integration lengths. Specifically, in a uniform system and for $L_x \gg L_y$, the average value of $C^2$ should behave as:

$$\langle C^2(L_x) \rangle \approx \frac{1}{L_x}\int_0^{L_x} dx\, \left[g_1(x,0)\right]^2 \propto \left(\frac{1}{L_x}\right)^{2\alpha} \qquad (1)$$

The long-range physics is then captured in a single parameter, the exponent $\alpha$ which describes the decay of $\langle C^2 \rangle$ with $L_x$. It is straightforward to understand the expected values of $\alpha$ in two simple limits. In a system with true long-range order, $g_1$ would be constant and the interference fringes would be perfectly straight. In this case $\alpha = 0$, corresponding to no decay of the contrast upon integration. In the opposite limit, if $g_1$ decays exponentially on a length scale much shorter than $L_x$, the integral in (1) is independent of $L_x$. In this case $\alpha = 0.5$, corresponding to adding up local interference fringes with random phases[14]. One of the central predictions of the BKT theory is that at the transition, the superfluid density should suddenly jump to a finite value which is a universal function of the transition temperature[29]. When adapted to the interference measurements with uniform 2D Bose gases[28], this "universal jump in superfluid density" corresponds to a sudden drop in $\alpha$ from 0.5 to 0.25.

In our experiments, integration along $y$ is automatically performed in absorption imaging, with $L_y \sim 10$ μm fixed by the size of the quasi-condensates. Our system is also not uniform along $x$ and the average local contrast $c_x = \langle c(x) \rangle$ decreases smoothly towards the edges of the quasi-condensate due to the increasing effects of thermal excitations. For comparison with theory, we consider the integrated contrast:

$$\tilde{C}(L_x) = \frac{1}{L_x} \left| \int_{-L_x/2}^{L_x/2} c(x) e^{i\varphi(x)} dx \right| .$$

This would exactly coincide with $C$ in a uniform system. We extract the exponent $\alpha$ using only the quasi-uniform region where $c_x > 0.5\, c_0$. Figure 3a shows examples of the measured $\langle \tilde{C}^2 \rangle$ as a function of $L_x$ at a low and a high temperature, along with the fits by a power-law decaying function.

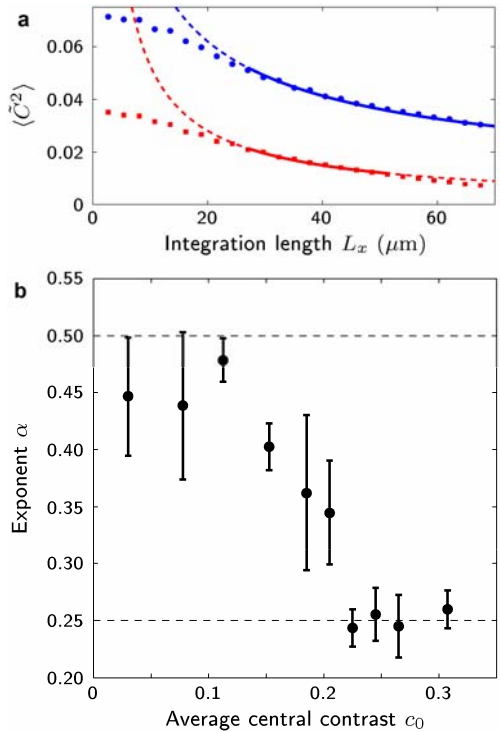

**Figure 3 | Emergence of quasi-long-range order in a 2D gas. a,** Examples of average integrated interference contrasts $\langle \tilde{C}^2(L_x) \rangle$ are shown for a low (blue circles, $c_0$=0.24) and a high temperature (red squares, $c_0$=0.13). The lines are fits to the data by the power law function $(1/L_x)^{2\alpha}$ and give $\alpha = 0.29 \pm 0.01$ (low temperature) and $\alpha = 0.46 \pm 0.01$ (high temperature). The fitting range, indicated by the solid part of the line, is constrained by the conditions $L_x \gg L_y$ on the left and $c_x > c_0/2$ on the right. **b,** Decay exponent $\alpha$ as a function of $c_0$. Dashed lines indicate the theoretically expected values of $\alpha$ above and below the BKT transition in a uniform system. Error bars indicate the standard deviation of the results from different experimental runs.

Figure 3b summarizes the fitted values of the exponent $\alpha$ in different temperature regimes and constitutes the first main result of this paper. Starting at high temperatures, for values of $c_0$ up to about 13%, $\alpha$ is approximately constant and close to 0.5.

When the temperature is reduced further, $\alpha$ rapidly drops to about 0.25, and for even lower temperatures (larger $c_0$) it levels off. We thus clearly observe a transition between two qualitatively different regimes at high and low temperatures. The values of $\alpha$ above and below the transition are in agreement with the theoretically expected jump in the superfluid density at the BKT transition in a uniform system. However, this quantitative agreement might be partly fortuitous. Even though we concentrated on the quasi-uniform part of the images, the geometry effects in our elongated samples could still be important. Ultimately at extremely low temperature, $\alpha$ should slowly tend to zero and the gas should become a pure, fully coherent BEC. We could not reach this regime in present experiments due to the residual heating discussed above.

Even without precise thermometry, we can estimate the cloud's temperature and density at the onset of quasi-long-range coherence. For images with $c_0$=0.15, the temperature inferred from the wings of the atom distribution after TOF is 290±40 nK, corresponding to a thermal wavelength of $\lambda$=0.3 μm. From the length of the quasi-condensate we deduce the number of condensed atoms $N_C$=11000±3000, and the peak condensate density (in the trap centre) $\rho_C$=(5±1)×$10^9$ cm$^{-2}$. This gives $\rho_C \lambda^2$=6±2. BKT theory for a uniform system predicts the transition at $\rho_S \lambda^2$=4, where $\rho_S$ is the superfluid density. The two values are in fair agreement, but we note that the exact relation between $\rho_C$ and $\rho_S$ in 2D atomic gases will require further experimental and theoretical investigation. For example, our observation of $\alpha \sim 0.5$ for a finite value of $c_0$ suggests that the superfluid density $\rho_S$ might be zero even if the condensate density $\rho_C$ is finite.

The key role in the microscopic BKT theory is played by vortices, localized topological defects in the phase of the condensate. In contrast to the smooth variation of the fringe phase $\varphi(x)$ created by long-wavelength phonons (fig. 1d), a free vortex in one of the condensates should appear as a sharp dislocation in the interference pattern[16,24], with $\varphi(x)$ changing abruptly across a dislocation line parallel to the expansion axis $z$. We indeed occasionally observe such dislocations. Examples of images containing one and several dislocations are shown in Figure 4a and 4b, respectively. The tightly bound vortex-antivortex pairs are not detectable in our experiments because they create only infinitesimal phase slips in the interference pattern. Other phase configurations which could mimic the appearance of a vortex, such as a dark soliton aligned with the imaging direction, can be discarded on theoretical grounds[24].

Figure 4c shows the frequency with which we detect sharp dislocations at different temperatures. For the count we consider only the central, 30 μm wide region of each image, which is smaller than the length of our smallest quasi-condensates. We note that we detect only a subset of vortices, i.e. those which are well isolated and close to the centre of the cloud. We also note that thermally activated phonon modes with a very short wavelength along $x$ can in principle contribute to the count. Their contribution is expected to be non-negligible only at the highest temperatures. There a detailed theoretical analysis would be needed to separate their effect from that of the vortices.

The observed sudden onset of vortex proliferation with increasing temperature constitutes the second main result of this paper. Further, this onset coincides with the loss of quasi-long-range coherence (Figure 3b). These two observations together provide conclusive evidence for the observation of the BKT crossover in this system.

Our experiments support the notion that the unbinding of vortex-antivortex pairs is the microscopic mechanism destroying the quasi-long-range coherence in 2D systems. The related question of the superfluidity of the sample remains open. It could be addressed in the future by setting the planar gases in rotation and studying the ordering of the vortex lattice. Alternatively, a study of the damping of the collective eigenmodes of the gas could be used to infer its viscosity. Our experiments may also open new theoretical questions related to the geometry and the mesoscopic nature of the system.

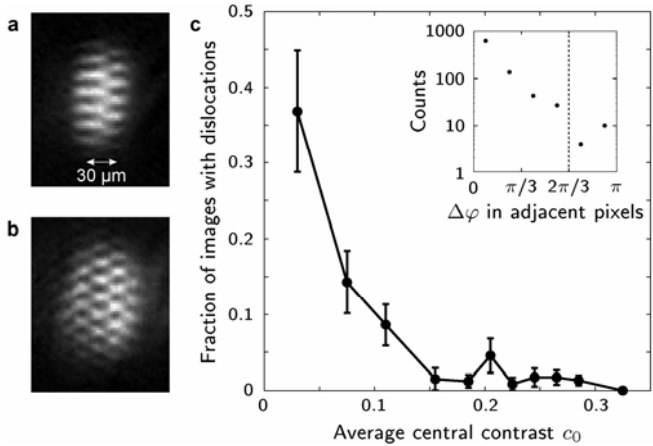

**Figure 4 | Proliferation of free vortices at high temperature**. **a,** Example of an interference pattern showing a sharp dislocation that we attribute to the presence of a free vortex in one of the interfering clouds. **b,** Interference pattern showing several dislocations. **c,** Fraction of images showing at least one dislocation in the central, 30 μm wide region, plotted as a function of $c_0$. The error bars show the statistical uncertainty, given by the square root of the number of images with dislocations. **Inset,** Histogram of the phase jumps $\Delta\varphi_i = |\varphi(x_i)-\varphi(x_{i+1})|$ between adjacent CCD pixel columns, for the set of images in the bin $c_0$=0.08. An image is counted as showing a dislocation if at least one of the $\Delta\varphi_i$ exceeds $2\pi/3$ (threshold indicated by the dashed line). The distance between adjacent columns is 2.7 μm and the count runs over the 10 central columns. There are 97 images contributing to this histogram, hence 970 counts, among which 16 counts (corresponding to 13 different images) exceed the threshold.

**Acknowledgments** We thank E. Altman, E. Demler, M. Lukin, A. Polkovnikov, P.-S. Rath, D. Stamper-Kurn, and S. Stock for useful discussions. We acknowledge financial support by IFRAF, ACI nanoscience, ANR, the Alexander von Humboldt foundation (P.K.) and the EU under Marie-Curie Fellowships (Z.H. and P.K.). Laboratoire Kastler Brossel is a research unit of Ecole normale supérieure and Université Paris 6, associated to CNRS.

**Author information** The authors declare that they have no competing financial interests Correspondence and request for materials should be addressed to J.D. (jean.dalibard@lkb.ens.fr).

1. Mermin, N. D. & Wagner, H. Absence of ferromagnetism or antiferromagnetism in one- or two-dimensional isotropic Heisenberg models. *Phys. Rev. Lett.* **17**, 1133-1136 (1966).
2. Hohenberg, P. C. Existence of long-range order in 1 and 2 dimensions. *Phys. Rev.* **158**, 383-386 (1967).
3. Berezinskii, V. L. Destruction of long-range order in one-dimensional and two-dimensional systems possessing a continuous symmetry group. II. Quantum systems. *Sov. Phys. JETP* **34**, 610-616 (1972).
4. Kosterlitz, J. M. & Thouless, D. J. Ordering, metastability and phase-transitions in 2 dimensional systems. *J. Phys. C* **6** 1181-1203 (1973).
5. Bishop, D.J. & Reppy, J. D. Study of the superfluid transition in two-dimensional $^4$He films. *Phys. Rev. Lett.* **40**, 1727-1730 (1978).
6. Resnick, D. J., Garland, J. C., Boyd, J. T., Shoemaker, S. & Newrock, R. S. Kosterlitz-Thouless transition in proximity-coupled superconducting arrays. *Phys. Rev. Lett.* **47**, 1542-1545 (1981).
7. Safonov, A.I., Vasilyev, S.A., Yasnikov, I.S., Lukashevich, I.I. & Jaakkola S. Observation of quasicondensate in two-dimensional atomic hydrogen. *Phys. Rev. Lett.* **81**, 4545-4548 (1998).
8. Görlitz, A. et al. Realization of Bose-Einstein condensates in lower dimensions. *Phys. Rev. Lett.* **87**, 130402 (2001).
9. Schweikhard, V., Coddington, I., Engels, P., Mogendorff, V. P. & Cornell, E. A. Rapidly rotating Bose-Einstein condensates in and near the lowest Landau level. *Phys. Rev. Lett.* **92**, 040404 (2004).
10. Rychtarik, D., Engeser, B., Nägerl, H.-C. & Grimm, R. Two-dimensional Bose-Einstein condensate in an optical surface trap. *Phys. Rev. Lett.* **92**, 173003 (2004).
11. Smith, N. L., Heathcote, W. H., Hechenblaikner, G., Nugent, E. & Foot, C. J. Quasi-2D confinement of a BEC in a combined optical and magnetic potential. *J. Phys. B* **38**, 223-235 (2005).
12. Orzel, C., Tuchman, A. K., Fenselau, M. L., Yasuda, M. & Kasevich, M. A. Squeezed states in a Bose-Einstein condensate. *Science* **291**, 2386-2389 (2001).
13. Burger, S. et al.Quasi-2D Bose-Einstein condensation in an optical lattice. *Europhys. Lett.* **57**, 1-6 (2002).
14. Hadzibabic, Z., Stock, S., Battelier, B., Bretin, V. & Dalibard, J. Interference of an array of independent Bose-Einstein condensates. *Phys. Rev. Lett.* **93**, 180403 (2004).
15. Köhl, M., Moritz, H., Stöferle, T., Schori, C. & Esslinger, T. Superfluid to Mott insulator transition in one, two, and three dimensions. *J. Low Temp. Phys.* **138**, 635-644 (2005).
16. Stock, S., Hadzibabic, Z., Battelier, B., Cheneau, M. & Dalibard, J. Observation of phase defects in quasi-two-dimensional Bose-Einstein condensates. *Phys. Rev. Lett.* **95**, 190403 (2005).
17. Bagnato, V. & Kleppner, D. Bose-Einstein condensation in low-dimensional traps. *Phys. Rev. A* **44**, 7439-7441 (1991).
18. Petrov, D.S., Holzmann, M. & Shlyapnikov, G. V. Bose-Einstein condensation in Quasi-2D Trapped Gases. *Phys. Rev. Lett.* **84**, 2551-2554 (2000).
19. Fernandez, J. P. & Mullin, W. J. The two-dimensional Bose-Einstein condensate. *J. Low Temp. Phys.* **128**, 233-249 (2002).
20. Andersen, J. O., Al Khawaja, U. & Stoof, H. T. C. Phase fluctuations in atomic Bose gases. *Phys. Rev. Lett.* **88**, 070407 (2002).
21. Petrov, D. S., Gangardt, D. M. & Shlyapnikov, G. V. Low-dimensional trapped gases. *J. Physique IV France* **116**, 5-44 (2004).
22. Simula, T. P., Lee, M. D. & Hutchinson, D. A. Transition from the Bose-Einstein condensate to the Berezinskii-Kosterlitz-Thouless phase. *Phil. Mag. Lett.* **85**, 395-403 (2005).
23. Holzmann, M., Baym, G., Blaizot, J.-P. & Laloë, F. The Kosterlitz-Thouless-Berezinskii transition of homogeneous and trapped Bose gases in two dimensions. Preprint cond-mat/0508131 at <http.arxiv.org> (2005).
24. Simula, T.P. & Blakie, P. B. Thermal activation of vortex-antivortex pairs in quasi-two-dimensional Bose-Einstein condensates. *Phys. Rev. Lett.* **96**, 020404 (2006).
25. Bramwell, S. T. & Holdsworth, P. C. W. Magnetization: A characteristic of the Kosterlitz-Thouless-Berezinskii transition. *Phys. Rev. B* **49**, 8811-8814 (1994).
26. Andrews, M. R. et al. Observation of interference between two Bose condensates. *Science* **275**, 637-641 (1997).
27. Hellweg, D. et al. Measurement of the Spatial Correlation Function of Phase Fluctuating Bose-Einstein Condensates. *Phys. Rev. Lett.* **91**, 010406 (2003).
28. Polkovnikov, A., Altman, E. & Demler, E. Interference between independent fluctuating condensates. *Proc. Natl. Acad. Sci. USA* **103**, 6125-6129 (2006).
29. Nelson, D. R. & Kosterlitz, J. M. Universal jump in superfluid density of 2-dimensional superfluids. *Phys. Rev. Lett.* **39**, 1201-1205 (1977).